\documentstyle[aps,12pt,preprint,psfig]{revtex}

\draft

\begin{document}
\tightenlines
\preprint{\hbox{PKU-TP-98-55}}

\title{Diffractive gluon jet production at
        hadron colliders in the two-gluon exchange model}

\author{Feng Yuan}
\address{\small {\it Department of Physics, Peking University, Beijing 100871, People's Republic
of China}}
\author{Kuang-Ta Chao}
\address{\small {\it China Center of Advanced Science and Technology (World Laboratory), Beijing 100080,
        People's Republic of China\\
      and Department of Physics, Peking University, Beijing 100871, People's Republic of China}}

\maketitle
\begin{abstract}

Following our recent paper on the calculations of diffractive quark
jet production at hadron colliders, we present here the calculations
of gluon jet production at hadron colliders in the two-gluon exchange
parameterization of the Pomeron model.
We use the helicity
amplitude method to calculate the cross section formula. We find that for the
gluon jet production the diffractive process is related to the
differential off-diagonal gluon distribution function in the proton.
We estimate the production rate for this process at the Fermilab Tevatron
by approximating the off-diagonal gluon distribution function by the usual
diagonal gluon distribution.
\end{abstract}


\section{Introduction}

In recent years, there has been a renaissance of interest in
diffractive scattering.
These diffractive processes are described by the Regge theory in
terms of the Pomeron ($I\!\! P$) exchange\cite{pomeron}.
The Pomeron carries quantum numbers of the vacuum, so it is a colorless entity
in QCD language, which may lead to the ``rapidity gap" events in experiments.
However, the nature of Pomeron and its reaction with hadrons remain a mystery.
For a long time it had been understood that the dynamics of the
``soft pomeron'' is deeply tied to confinement.
However, it has been realized now that how much can be learned about
QCD from the wide variety of small-$x$ and hard diffractive processes,
which are now under study experimentally.
In Refs.\cite{th1,th2}, the diffractive $J/\psi$ and $\Upsilon$ production
cross section have been formulated in photoproduction processes and in
DIS processes in perturbative QCD.
In the framework of perturbative QCD the Pomeron is represented by a pair of
gluon in the color-singlet sate.
This two-gluon exchange model can successfully describe the experimental
results from HERA\cite{hera-ex}.

On the other hand, as we know that there exist nonfactorization effects
in the hard diffractive processes at hadron colliders
\cite{preqcd,collins,soper,tev}.
First, there is the so-called spectator effect\cite{soper}, which can
change the probability of the diffractive hadron emerging from collisions
intact. Practically, a suppression factor (or survive factor) ``$S_F$"
is used to describe this effect\cite{survive}.
Obviously, this suppression factor can not be calculated in perturbative
QCD, which is now viewed as a nonperturbative parameter.
Typically, the suppression factor $S_F$ is determined to be about
$0.1$ at the energy scale of the Fermilab Tevatron\cite{tev}.
Another nonfactorization effect discussed in literature is associated with the coherent
diffractive processes at hadron colliders\cite{collins}, in which
the whole Pomeron is induced in the hard scattering.
It is proved in \cite{collins} that the existence of the leading twist
coherent diffractive processes is associated with a breakdown of the
QCD factorization theorem.

Based on the success of the two-gluon exchange parametrization of the Pomeron
model in the description of the diffractive photoproduction processes
at $ep$ colliders\cite{th1,th2,hera-ex}, we may extend the applications
of this model to calculate the diffractive processes at hadron colliders in perturbative QCD.
Under this context, the Pomeron represented by a color-singlet two-gluon
system emits from one hadron and interacts with another hadron in hard
process, in which the two gluons are both involved (as shown in Fig.~1).
Therefore, these processes calculated in the two-gluon exchange model are just
belong to the coherent diffractive processes in hadron collisions.
Another important feature of the calculations of the diffractive processes in this model recently demonstrated is the sensitivity
to the off-diagonal parton distribution
function in the proton\cite{offd}.

Using this two-gluon exchange model, we have calculated the  diffractive
$J/\psi$ production \cite{psi}, quark jet production\cite{charm,quark},
massive muon pair and $W$ boson productions\cite{dy},
and direct photon production\cite{photon} in hadron collisions.
In this paper, we will further calculate the gluon jet production
at large transverse momentum in the coherent diffractive processes at hadron
colliders by using the two-gluon exchange model.
In the calculations of Refs.\cite{psi,charm,dy}, there always is a large mass
scale associated with the production process.
That is $M_\psi$ for $J/\psi$ production, $m_c$ for the charm jet
production, $M^2$ for the massive muon production ($M^2$ is the invariant mass
of the muon pair) and $M_W^2$ for $W$ boson production.
However, in the gluon jet production process as well as the light quark jet production process, 
there is no large mass scale.
So, for these processes, the large transverse momentum is needed to
guarantee the application of the perturbative QCD.
Furthermore, in \cite{quark} we
 show that the light quark jet production
in the two-gluon exchange model has a distinctive feature that
there is no contribution from the small $l_T^2$ region ($l_T^2<k_T^2$)
in the integration of the amplitude over $l_T^2$. (The similar behavior
has also been found for the diffractive light quark photoproduction
process\cite{zaka}.)
So, the expansion (in terms of $l_T^2/M_X^2$) method 
used in Refs.\cite{psi,charm,dy} can not be applied to the calculations
of gluon jet production.
In the following calculations, we will employ the helicity amplitude method
to calculate the amplitude of the diffractive gluon jet production
in hadron collisions.
We will show that the production
cross section is related to the differential (off-diagonal) gluon distribution
function in the proton as that in the diffractive light quark jet
production process\cite{quark}.
(On the other hand, we note that the cross sections of the processes
calculated in Refs.\cite{psi,charm,dy} are related to 
the integrated gluon distribution function in the proton).

Diffractive gluon jet production can come from two types partonic processes:
one is the quark initiated process (Fig.2), and the other is the gluon initiated process
(Fig.3).

The diffractive production of heavy quark jet at hadron colliders has also
been studied by using the two-gluon exchange model in Ref.\cite{levin}.
However, their calculation method is very different from ours
\footnote{For detailed discussions and comments, please see \cite{charm}}.
In their calculations, they separated their diagrams into two parts,
and called one part the coherent diffractive contribution to the heavy
quark production.
However, this separation can not guarantee the
gauge invariance\cite{charm}.
In our approach, we follow the definition of
Ref.\cite{collins}, i.e., we call
the process in which the whole Pomeron participants in the hard scattering
process as the coherent diffractive process.
Under this definition, all of the diagrams plotted in Fig.2
and Fig.3 for the partonic processes contribute to
the coherent diffractive production.

The rest of the paper is organized as follows.
In Sec.II, we will give the cross section formula for the partonic process
in the leading order of perturbative QCD.
In this section we employ the helicity amplitude method to calculate
two partonic processes, $qp\rightarrow qgp$ and $gp\rightarrow ggp$.
In Sec.III, we estimate the production rate of diffractive gluon
jet at the Fermilab Tevatron by approximating the off-diagonal
gluon distribution function by the usual diagonal gluon distribution
function in the proton.
We also compare the contributions from different partonic processes to
the diffractive dijet production at the Tevatron.
And the conclusions will be given in Sec.IV.

\section{ The cross section formula for the partonic process}

\subsection{$qp\rightarrow qgp$ process}
For the partonic process $qp\rightarrow qgp$, in the leading order of perturbative
QCD, there are nine diagrams shown in Fig.2.
The two-gluon system coupled to the proton (antiproton) in Fig.2 is in
a color-singlet state, which characterizes the diffractive processes in
perturbative QCD.
Due to the positive signature of these diagrams (color-singlet exchange),
we know that the real part of the amplitude cancels out in the leading
logarithmic approximation.
To get the imaginary part of the amplitude, we must calculate the
discontinuity represented by the crosses in each diagram of Fig.2.

The first four diagrams of Fig.2 are the same as those calculated in the
diffractive direct photon production process at hadron colliders\cite{photon}.
But, due to the existence of gluon-gluon interaction vertex in QCD,
in the partonic process $qp\rightarrow qgp$, there are additional five diagrams (Fig.2(5)-(9)).
These five diagrams are needed for a complete calculation in this order of QCD.

In our calculations, we express the formulas in terms of the Sudakov
variables.
That is, every four-momenta $k_i$ are decomposed as,
\begin{equation}
k_i=\alpha_i q+\beta_i p+\vec{k}_{iT},
\end{equation}
where $q$ and $p$ are the momenta of the incident quark and the proton,
$q^2=0$, $p^2=0$, and $2p\cdot q=W^2=s$.
Here $s$ is the c.m. energy of the quark-proton system, i.e., the invariant
mass of the partonic process $qp\rightarrow qg p$.
$\alpha_i$ and $\beta_i$ are the momentum fractions of $q$ and $p$
respectively.
$k_{iT}$ is the transverse momentum, which satisfies
\begin{equation}
k_{iT}\cdot q=0,~~~
k_{iT}\cdot p=0.
\end{equation}

All of the Sudakov variables for every momentum
are determined by using the on-shell conditions
of the momenta represented by the external lines and the crossed lines in the diagram.
The calculations of these Sudakov variables are similar to those in
the diffractive light quark jet production process $gp\rightarrow q\bar q p$\cite{quark},
and we can get the Sudakov variables of every momentum for the process
$qp\rightarrow qgp$ from the relevant formulas of \cite{quark}.
In the following, we list all of the Sudakov variables for the diffractive
process $qp\rightarrow qgp$.

For the momentum $u$, we have
\begin{equation}
\alpha_u=0,~~\beta_u=x_{I\! P}=\frac{M_X^2}{s},~~u_T^2=t=0,
\end{equation}
where $M_X^2$ is the invariant mass squared of the diffractive final state
including the light quark and antiquark jets.
For the high energy diffractive process, we know that $M_x^2\ll s$, so
we have $\beta _u$ ($x_{I\! P}$) as a small parameter.
For the momentum $k$,
\begin{equation}
\label{ak}
\alpha_k(1+\alpha_k)=-\frac{k_T^2}{M_X^2},~~\beta_k=-\alpha_k\beta_u,
\end{equation}
where $k_T$ is the transverse momentum of the out going quark jet.
For the loop mentum $l$, because the results for $\beta_l$ are not the
same for the nine diagrams of Fig.2, we get its value from the formula
of Ref.\cite{quark} for the relevant diagram. The results are
\begin{eqnarray}
\nonumber
\alpha_l&=&-\frac{l_T^2}{s},\\
\nonumber
\beta_l&=&\frac{2(k_T,l_T)-l_T^2}{\alpha_ks},~~~{\rm for~Diag.}1,~2,~6,\\
\nonumber
&=&\frac{2(k_T,l_T)+l_T^2}{(1+\alpha_k)s},~~~{\rm for~Diag.}5,~7,~8,\\
&=&-\frac{M_X^2-l_T^2}{s},~~~~~~~{\rm for~Diag.}3,~4,~9,
\end{eqnarray}
where $(k_T,l_T)$ is the 2-dimensional product of the transverse vectors
$\vec{k}_T$ and $\vec{l}_T$.

Using these Sudakov variables, we can give the cross section formula
for the partonic process $qp\rightarrow qgp$ as,
\begin{equation}
\label{xs}
\frac{d\hat{\sigma}(qp\rightarrow qgp)}{dt}|_{t=0}=\frac{dM_X^2d^2k_Td\alpha_k}{16\pi s^216\pi^3M_X^2}
        \delta(\alpha_k(1+\alpha_k)+\frac{k_T^2}{M_X^2})\sum \overline{|{\cal A}|}^2,
\end{equation}
where ${\cal A}$ is the amplitude of the process $qp\rightarrow qgp$.
We know that the real part of the amplitude ${\cal A}$ is zero,
and the imaginary part of the amplitude ${\cal A}(qp\rightarrow qgp)$ for each diagram
of Fig.2 has the following general form,
\begin{equation}
\label{ima}
{\rm Im}{\cal A}=C_F(T_{ij}^a)\int \frac{d^2l_T}{(l_T^2)^2}F\times\bar u
        _i(u-k)\Gamma_\mu u_j(q),
\end{equation}
where $C_F$ is the color factor for each diagram.
$a$ is the color index of the incident gluon.
$\Gamma_\mu$ represents some $\gamma$ matrices including one
propagator. $F$ in the integral is defined as
\begin{equation}
\label{feq}
F=\frac{3}{2s}g_s^3f(x',x^{\prime\prime};l_T^2),
\end{equation}
where
\begin{equation}
\label{offd1}
f(x',x^{\prime\prime};l_T^2)=\frac{\partial G(x',x^{\prime\prime};l_T^2)}{\partial {\rm ln} l_T^2},
\end{equation}
where the function
$G(x',x^{\prime\prime};k_T^2)$ is the so-called
off-diagonal gluon distribution function\cite{offd}.
Here, $x'$ and $x^{\prime\prime}$ are the momentum fractions of the proton
carried by the two gluons.
It is expected that at small $x$, there is no big difference between the off-diagonal and
the usual diagonal gluon densities\cite{off-diag}.
So, in the following calculations, we estimate the production rate by
approximating the off-diagonal gluon density by 
the usual diagonal gluon density, 
$G(x',x^{\prime\prime};Q^2)\approx xg(x,Q^2)$, where $x=x_{I\!\! P}=M_X^2/s$.

In \cite{photon}, we calculate the cross section for the diffractive photon
production process $qp\rightarrow \gamma qp$ by directly squaring the
partonic process amplitude.
However, in the calculations here for the partonic process $qp\rightarrow qgp$
because there are additional five diagrams contribution, it is not convenient
to directly square the amplitude.
Following Ref.\cite{quark}, we calculate the amplitude
by employing the {\it helicity amplitude} method\cite{ham,wu}.
Furthermore, we will show that by using the helicity amplitude method we can
reproduce the cross section formula for the diffractive photon production
process\cite{photon}.

For the massless quark spinors, we define
\begin{equation}
u_\pm(p)=\frac{1}{\sqrt{2}}(1\pm\gamma_5)u(p).
\end{equation}
For the polarization vector of the outgoing gluon (its momentum is $k+q$), 
following the method of Ref.\cite{wu}, we find that it is convenient to
choose
\begin{equation}
\label{epq}
\not\! e^{(\pm)}=N_e[(\not\! k+\not\! q)\not\! q\not\! p(1\mp\gamma_5)+
        \not\! p\not\! q(\not\! k+\not\! q)(1\pm\gamma_5)].
\end{equation}
The normalization factor $N_e$ equals to
\begin{equation}
\label{ene}
N_e=\frac{1}{s\sqrt{2k_T^2}}.
\end{equation}
With this definition (\ref{epq}), we can easily get the scalar products between the
four-momenta and the polarization vector $e$ as
\begin{equation}
\label{eprc}
e\cdot p=0,~~e\cdot q=N_e\frac{k_T^2s}{1+\alpha_k},~~e\cdot k_T=-N_ek_T^2s,~~e\cdot l_T=-N_e(k_T,l_T)s.
\end{equation}

The helicity amplitudes for the processes in which the polarized Dirac particles are involved
have the following general forms\cite{ham},
\begin{equation}
\label{ham}
\bar u_\pm(p_f)Qu_\pm(p_i)=\frac{Tr[Q\not\! p_i\not\! n\not\! p_f(1\mp\gamma_5)]}
        {4\sqrt{(n\cdot p_i)(n\cdot p_f)}},
\end{equation}
where $n$ is an arbitrary massless 4-vector, which is set to be $n=p$ in the
following calculations.
Using this formula (\ref{ham}), the calculations of the helicity amplitude
${\cal A}(\lambda_1,\lambda_2,\lambda_3)$ for the diffractive
process $qp\rightarrow qgp$ is straightforward.
Here $\lambda_1$ represents the helicity of the incident quark;
$\lambda_2$ and $\lambda_3$ represent the helicities of the
outgoing gluon and quark respectively.
In our calculations, we only take the leading order contributions, and neglect the
higher order contributions which are proportional to $\beta_u=\frac{M_X^2}{s}$
because in the high energy diffractive processes we have $\beta_u\ll 1$.

For the first four diagrams, to sum up together, the imaginary part of the
amplitude ${\cal A}(+,+,+)$ is
\begin{equation}
\label{im1}
{\rm Im}{\cal A}^{1234}(+,+,+)=\alpha_k^2(1+\alpha_k){\cal N}\times
        \int\frac{d^2\vec{l}_T}{(l_T^2)^2}f(x',x'';l_T^2)
        (\frac{2}{9}-\frac{-1}{36}\frac{k_T^2-(1+\alpha_k)(k_T,l_T)}{(\vec{k}_T-(1+\alpha_k)\vec{l}_T)^2}),
\end{equation}
where $\frac{2}{9}$ and $\frac{-1}{36}$ are the color factors for Diags.1,4 and
Diags.2,3 respectively, and
${\cal N}$ is defined as
\begin{equation}
{\cal N}=\frac{3s}{\sqrt{-2\alpha_kk_T^2}}g_s^3T_{ij}^a.
\end{equation}
The other helicity amplitudes for the first four diagrams have the similar forms
as (\ref{im1}),
\begin{eqnarray}
\label{im11}
\nonumber
{\rm Im}{\cal A}^{1234}(-,-,-)&=&{\rm Im}{\cal A}^{1234}(+,+,+),\\
{\rm Im}{\cal A}^{1234}(+,-,+)&=&{\rm Im}{\cal A}^{1234}(-,+,-)=\frac{-1}{\alpha_k}{\rm Im}{\cal A}^{1234}(+,+,+).
\end{eqnarray}

These amplitude expressions Eq.~(\ref{im1}) can also serve as the calculations
of the amplitude for the diffractive direct photon
production process $q p\rightarrow q\gamma p$ \cite{photon} except the difference
on the color factors.\footnote{In Ref.\cite{photon}, we did not employ the
helicity amplitude method. If we use the amplitude expressions Eqs.(\ref{im1}) and
(\ref{im11}) (correct the color factors) to calculate the photon production process $qp\rightarrow q\gamma p$,
we can get the same result as that in \cite{photon}. This can be viewed as a cross
check for the methods we used in the calculations.}
In the direct photon process, the color
factors for these four diagrams are the same (they are all $\frac{2}{9}$).
It is instructive to see what is the consequence of this difference.
We know that the amplitude of the diffractive process in Eq.~(\ref{ima}) must be
zero in the limit $l_T^2\rightarrow 0$. Otherwise, this will lead to a linear singularity
when we perform the integration of the amplitude over $l_T^2$ due to existence of the factor $1/(l_T^2)^2$ in the
integral of Eq.~(\ref{ima})\cite{charm}.
This linear singularity is not proper in QCD calculations.
So, we must first exam the amplitude behavior under the limit of $l_T^2\rightarrow 0$
for all the diffractive processes in the calculations using the two-gluon exchange model.
From Eq.~(\ref{im1}), we can see that the amplitude for the
diffractive direct photon production process $qp\rightarrow q\gamma p$ is exact zero at $l_T^2\rightarrow 0$.
However, for the process $qp\rightarrow qgp$ the amplitude
for the first four diagrams 
is not exact zero in the limit $l_T^2\rightarrow 0$ due to the inequality of the
color factors between them.
So, for this process there must be other diagrams in this order
of perturbative QCD calculation to cancel out the linear singularity which
rises from the first four diagrams.
The last five diagrams of Fig.2 are just for this purpose.

Finally, by adding up all of the nine diagrams of Fig.2, the 
imaginary parts of the amplitudes
are
\begin{eqnarray}
\nonumber
{\rm Im}{\cal A}(+,+,+)&=&{\rm Im}{\cal A}(-,-,-)=\frac{\alpha_k^2}{4}{\cal N}\times {\cal T},\\
{\rm Im}{\cal A}(+,-,+)&=&{\rm Im}{\cal A}(-,+,-)=-\frac{\alpha_k}{4}{\cal N}\times {\cal T},
\end{eqnarray}
where
\begin{eqnarray}
\label{int2}
\nonumber
{\cal T}&=&\int\frac{d^2\vec{l}_T}{(l_T^2)^2}f(x',x'';l_T^2)[
        \frac{(1+\alpha_k)^2}{9}\frac{(k_T,l_T)-(1+\alpha_k)l_T^2}{(\vec{k}_T-(1+\alpha_k)\vec{l}_T)^2}
        -(1+\alpha_k)\frac{(k_T,l_T)+l_T^2}{(\vec{k}_T+\vec{l}_T)^2}\\
  &&-\alpha_k\frac{(k_T,l_T)-l_T^2}{(\vec{k}_T-\vec{l}_T)^2}
  +\alpha_k^2\frac{(k_T,l_T)-\alpha_kl_T^2}{(\vec{k}_T-\alpha_k\vec{l}_T)^2}].
\end{eqnarray}
From the above results, we can see that in the integration of the amplitude
the linear singularity from different diagrams are canceled out by each other,
which will guarantee there is no linear singularity in the total sum.

Another feature of the above results for the amplitudes is the relation to the
differential off-diagonal gluon distribution function $f(x',x'';l_T^2)$.
However, as mentioned above that there is no big difference between the off-diagonal
gluon distribution function and the usual gluon distribution at small $x$,
so we can simplify the integration of (\ref{int2}) by 
approximating the differential off-diagonal gluon
distribution function $f(x',x'';l_T^2)$ by the usual diagonal differential
gluon distribution function $f_g(x;l_T^2)$.

After integrating over the azimuth angle of $\vec{l}_T$, the integration
${\cal T}$ will then be
\begin{eqnarray}
\label{qt}
\nonumber
{\cal T}&=&\pi\int\frac{dl_T^2}{(l_T^2)^2}f_g(x;l_T^2)[
        \frac{1+\alpha_k}{9}(\frac{1}{2}-\frac{k_T^2-(1+\alpha_k)l_T^2}{2|k_T^2-(1+\alpha_k)l_T^2|})
        +(\frac{1}{2}-\frac{k_T^2-l_T^2}{2|k_T^2-l_T^2|})\\
        &&+\alpha_k(\frac{1}{2}-\frac{k_T^2-\alpha_k l_T^2}{2|k_T^2-\alpha_k l_T^2|})].
\end{eqnarray}
In the above integration, if $l_t^2<k_T^2/(1+\alpha_k)^2$ the first term of the
integration over $l_T^2$ will be zero; if $l_t^2<k_T^2$ the second
term will be zero; if $l_t^2<k_T^2/\alpha_k^2$ the third term will be zero.
So, the dominant regions contributing to the three integration terms are
$l_t^2\sim k_T^2/(1+\alpha_k)^2$, $l_t^2\sim k_T^2$, and $l_t^2\sim k_T^2/\alpha_k^2$
respectively.
Approximately, by ignoring some evolution effects of the differential gluon
distribution function $f_g(x;l_T^2)$ in the above dominant integration regions,
we get the following results for the integration ${\cal T}$,
\begin{equation}
\label{i1}
{\cal T}=\frac{\pi}{k_T^2}[f_g(x;k_T^2)+\frac{(1+\alpha_k)^3}{9}f_g(x;\frac{k_T^2}{(1+\alpha_k)^2})
        +\alpha_k^3f_g(x;\frac{k_T^2}{\alpha_k^2})].
\end{equation}

Obtained the formula for the integration ${\cal T}$, the amplitude squared for
the partonic process $qp\rightarrow qgp$ will be reduced to, after averaging
over the spin and color degrees of freedom,
\begin{equation}
\overline{|{\cal A}|}^2=\frac{\alpha_s^3(4\pi)^3}{24}\frac{1+\alpha_k^2}{M_X^2(1+\alpha_k)}s^2|{\cal T}|^2.
\end{equation}
And the cross section for the partonic process $qp\rightarrow qgp$ is
\begin{eqnarray}
\label{xsp}
\nonumber
\frac{d\hat\sigma(qp\rightarrow qgp)}{dt}|_{t=0}&=&\int_{M_X^4>4k_T^2}dM_X^2dk_T^2d\alpha_k[\delta(\alpha_k-\alpha_1)+\delta(\alpha_k-\alpha_2)]\\
        &&\frac{\alpha_s^3}{96(M_X^2)^2}\frac{1+\alpha_K^2}{1+\alpha_k}\frac{1}{\sqrt{1-\frac{4k_T^2}{M_X^2}}}
        |{\cal T}|^2,
\end{eqnarray}
where $\alpha_{1,2}$ are the solutions of the following equations,
\begin{equation}
\alpha(1+\alpha)+\frac{k_T^2}{M_X^2}=0.
\end{equation}
The integral bound $M_X^2>4k_T^2$ in (\ref{xsp}) shows that the dominant contribution
of the integration over $M_X^2$ comes from the region of $M_X^2\sim 4k_T^2$.
Using Eq.~(\ref{ak}), this indicates that in this dominant region 
$\alpha_k$ is of order of 1.
So, in the integration ${\cal T}$ the differential gluon distribution
function $f_g(x;Q^2)$ of the three terms can approximately take their values
at the same scale of $Q^2=k_T^2$.
That is, the integration ${\cal T}$ is then simplified to
\begin{equation}
\label{i2}
{\cal T}=\frac{\pi}{9k_T^2}f_g(x;k_T^2)(1+\alpha_k)(10-7\alpha_k+10\alpha_k^2).
\end{equation}
Numerical calculations show that there is little difference
between the cross sections by using these two different
parametrizations of ${\cal T}$, Eq.~(\ref{i1}) and Eq.~(\ref{i2}).
So, in Sec.IV, we use Eqs.~(\ref{xsp}) and (\ref{i2}) to
estimate the diffractive production rate at the Fermilab
Tevatron.

\subsection{$gp\rightarrow ggp$ process}

For the partonic process $gp\rightarrow ggp$, there are twelve diagrams in the
leading order contributions as shown in Fig.3.
The first nine diagrams are due to the existence of the three-gluon interaction vertex, and
the last three diagrams are due to the existence of  the four-gluon interaction vertex.
But it will be shown in the following calculations, the last three diagrams
do not contribute under some choice of the polarizations of the three
external gluons.

The Sudakov variables can be calculated by the similar method used in the
last subsection. And those Sudakov variables of the momenta $u$, $k$ for the
process $gp\rightarrow ggp$ are the same as those in the last subsection.
For the loop momentum $l$, the relevant Sudakov variables for each diagram
are
\begin{eqnarray}
\nonumber
\alpha_l&=&-\frac{l_T^2}{s},\\
\nonumber
\beta_l&=&\frac{2(k_T,l_T)-l_T^2}{\alpha_ks},~~~{\rm for~Diag.}1,~4,~6,~10,\\
\nonumber
&=&\frac{2(k_T,l_T)+l_T^2}{(1+\alpha_k)s},~~~{\rm for~Diag.}2,~3,~5,~11,\\
&=&-\frac{M_X^2-l_T^2}{s},~~~~~~~{\rm for~Diag.}7,~8,~9,~12.
\end{eqnarray}

And also, we can express the cross section formula for the partonic process
$gp\rightarrow ggp$ in the following form,
\begin{equation}
\label{gxs}
\frac{d\hat{\sigma}(gp\rightarrow ggp)}{dt}|_{t=0}=\frac{dM_X^2d^2k_Td\alpha_k}{16\pi s^216\pi^3M_X^2}
        \delta(\alpha_k(1+\alpha_k)+\frac{k_T^2}{M_X^2})\sum \overline{|{\cal A}|}^2,
\end{equation}
where ${\cal A}$ is the amplitude of the process $gp\rightarrow ggp$.
We know that the real part of the amplitude ${\cal A}$ is zero,
and the imaginary part of the amplitude ${\cal A}(gp\rightarrow ggp)$ for each diagram
of Fig.3 has the following general form,
\begin{equation}
\label{gima}
{\rm Im}{\cal A}=C_Ff_{abc}\int \frac{d^2l_T}{(l_T^2)^2}G(e_1,e_2,e_3)\times F,
\end{equation}
where $C_F$ is the color factor for each diagram.
$a,~b,~c$ are the color indexes for the incident gluon and the two outgoing
gluons respectively, and $f_{abc}$ are the antisymmetric $SU(3)$ structure
constants.
$G(e_1,e_2,e_3)$ represents the interaction part including one propagator
for the first nine diagrams, where
$e_1,~e_2,~e_3$ are the polarization vectors for the incident gluon
and the two outgoing gluons.
$F$ in the integral is the same as that in Eq.(\ref{feq}).

The color factors $C_F$ for the twelve diagrams are 
\begin{eqnarray}
\label{cf}
\nonumber
C_F&=&\frac{1}{2},~~~~~~~{\rm for~ Diag.}1,~7,\\
\nonumber
C_F&=&-\frac{1}{2},~~~~~{\rm for~ Diag.}2,\\
\nonumber
C_F&=&-\frac{1}{4},~~~~~{\rm for~ Diag.}3,~6,~9,\\
\nonumber
C_F&=&\frac{1}{4},~~~~~~~{\rm for~ Diag.}4,~5,~8,\\
C_F&=&\frac{3}{4},~~~~~~~{\rm for~ Diag.}10,~11,~12.
\end{eqnarray}

Following the calculation method used in the last subsection, we employ the helicity amplitude method
to calculate the amplitude Eq.(\ref{gima}).
For the polarization vector of the incident gluon, which is transversely polarized,
we choose,
\begin{equation}
\label{ev}
e^{(\pm)}_1=\frac{1}{\sqrt{2}}(0,1,\pm i,0).
\end{equation}
For the two outgoing gluons, we choose their polarization vectors as\cite{wu}
\begin{eqnarray}
\label{geprc}
\nonumber
\not\! e_2^{(\pm)}=N_e[(\not\! k+\not\! q)\not\! q\not\! p(1\mp\gamma_5)+
        \not\! p\not\! q(\not\! k+\not\! q)(1\pm\gamma_5)],\\
\not\! e_3^{(\pm)}=N_e[(\not\! u-\not\! k)\not\! q\not\! p(1\mp\gamma_5)+
        \not\! p\not\! q(\not\! u-\not\! k)(1\pm\gamma_5)].
\end{eqnarray}
The normalization factor $N_e$ has the same form as in Eq.(\ref{ene}).
Under the above choice of the polarization vectors for the external gluons,
we can easily find that they are satisfied the following equations,
\begin{equation}
p\cdot e_1=p\cdot e_2=p\cdot e_3=0.
\end{equation}
With these relations, we can further find that the last three diagrams do
not contribute to the partonic process $gp\rightarrow ggp$.

For the first nine diagrams, there are two helicity amplitudes among
the eight helicity amplitudes do not contribute in the context of the above choice
of the polarizations of the external gluons, i.e.,
\begin{equation}
{\rm Im}{\cal A}(+,+,+)={\rm Im}{\cal A}(-,-,-)=0.
\end{equation}
In the expression of the amplitude ${\cal A}(\lambda(e_1),\lambda(e_2),\lambda(e_3))$,
$\lambda$ denote the helicities for the three gluons respectively.
The other six helicity amplitudes are divided into the following three different
sets,
\begin{eqnarray}
\nonumber
{\rm Im}{\cal A}(+,-,-)&\sim &{\rm Im}{\cal A}(-,+,+),\\
\nonumber
{\rm Im}{\cal A}(+,-,+)&\sim &{\rm Im}{\cal A}(-,+,-),\\
{\rm Im}{\cal A}(+,+,-)&\sim &{\rm Im}{\cal A}(-,-,+).
\end{eqnarray}

For the first helicity amplitudes set, ${\rm Im}{\cal A}(\pm,\mp,\mp)$,
to sum up all of the nine diagrams, we get
\begin{equation}
\label{gm1}
{\rm Im}{\cal A}(\pm,\mp,\mp)={\cal N}'\pi \vec{e}_1^{(\pm)}\cdot k_T{\cal I},
\end{equation}
where ${\cal N}'$ is defined as
\begin{equation}
{\cal N}'=\frac{3}{4}\frac{s}{k_T^2}g_s^3f_{abc}.
\end{equation}
And the integration ${\cal I}$ is
\begin{eqnarray}
\label{gim11}
\nonumber
{\cal I}&=&{1\over \pi}\int\frac{d^2\vec{l}_T}{(l_T^2)^2}f(x',x'';l_T^2)[-(1+\alpha_k)\frac{k_T^2+(k_T,l_T)}{(\vec{k}_T+\vec{l}_T)^2}
        +\alpha_k\frac{k_T^2-(k_T,l_T)}{(\vec{k}_T-\vec{l}_T)^2}
        \\
  &&+(1+\alpha_k)^2\frac{k_T^2-(1+\alpha_k)(k_T,l_T)}{(\vec{k}_T-(1+\alpha_k)\vec{l}_T)^2}
  -\alpha_k^2\frac{k_T^2-\alpha_k(k_T,l_T)}{(\vec{k}_T-\alpha_k\vec{l}_T)^2}]\\
\nonumber
&=&{1\over \pi}\int\frac{d^2\vec{l}_T}{(l_T^2)^2}f(x',x'';l_T^2)[(1+\alpha_k)\frac{(k_T,l_T)+l_T^2}{(\vec{k}_T+\vec{l}_T)^2}
        +\alpha_k\frac{(k_T,l_T)-l_T^2}{(\vec{k}_T-\vec{l}_T)^2}        
        \\
  &&-(1+\alpha_k)^2\frac{(k_T,l_T)-(1+\alpha_k)l_T^2}{(\vec{k}_T-(1+\alpha_k)\vec{l}_T)^2}
  +\alpha_k^2\frac{(k_T,l_T)-\alpha_kl_T^2}{(\vec{k}_T-\alpha_k\vec{l}_T)^2}].
\end{eqnarray}
From the above equations, we can check that there is no linear singularity at the
limit of $l_T^2\rightarrow 0$ in the integration of the amplitude over
the loop momentum.
The first term of the integration ${\cal I}$ in Eq.(\ref{gim11}) comes from the
contribution of Diag.3; the second term comes from Diag.4; the third term comes from Diag.6 and Diag.9;
the last term comes from Diag.5 and Diag.8. The contributions from Diags.1, 2
and 7 are canceled out by each other.
From (\ref{gim11}), we can see that the linear singularities coming from the
four terms are canceled out by each other.
The final result for the amplitude is now free of linear singularity.
We must emphasize here that only the total sum of the contributions from  all
of the diagrams is free of linear singularity. The separation of these diagrams
will cause linear singularity.

Following the argument in the last subsection of the calculation for the partonic
process $qp\rightarrow qgp$, we can approximate the differential off-diagonal gluon
distribution function $f(x',x'';l_T^2)$ by the usual diagonal differential
gluon distribution function $f_g(x;l_T^2)$ to further simplify the integration
of ${\cal I}$.
After integrating over the azimuth angle of $\vec{l}_T$, this integration
will then be
\begin{eqnarray}
\label{gi2}
\nonumber
{\cal I}&=&\int\frac{dl_T^2}{(l_T^2)^2}f_g(x;l_T^2)[(\frac{1}{2}-\frac{k_T^2-l_T^2}{2|k_T^2-l_T^2|})
        -{(1+\alpha_k)}(\frac{1}{2}-\frac{k_T^2-(1+\alpha_k)l_T^2}{2|k_T^2-(1+\alpha_k)l_T^2|})
        \\
        &&+\alpha_k(\frac{1}{2}-\frac{k_T^2-\alpha_k l_T^2}{2|k_T^2-\alpha_k l_T^2|})].
\end{eqnarray}
The above equation shows that the integration ${\cal I}$ here has the similar
behavior as that of the integration ${\cal T}$ of Eq.(\ref{qt}) in the last
subsection.
So, the three terms of the above integration ${\cal I}$ are dominantly 
contributed from the integral regions of $l_T^2$ as
$l_t^2\sim k_T^2/(1+\alpha_k)^2$, $l_t^2\sim k_T^2$, and $l_t^2\sim k_T^2/\alpha_k^2$
respectively.
Approximately, we may also ignore the evolution effects of the differential gluon
distribution function $f_g(x;l_T^2)$ in the above dominant integration regions,
and so the integration ${\cal I}$ is reduced to
\begin{equation}
\label{gi1}
{\cal I}=\frac{1}{k_T^2}[f_g(x;k_T^2)-(1+\alpha_k)^3f_g(x;\frac{k_T^2}{(1+\alpha_k)^2})
        +\alpha_k^3f_g(x;\frac{k_T^2}{\alpha_k^2})].
\end{equation}

For the second helicity amplitudes set, ${\rm Im}{\cal A}(\pm,\mp,\pm)$,
the calculations are more complicated, and
the contribution from Diag.3 is
\begin{eqnarray}
\label{g21}
\nonumber
{\rm Im}{\cal A}^3(\pm,\mp,\pm)&=&-{\cal N}'(1+\alpha_k)^2\int\frac{d^2\vec{l}_T}{(l_T^2)^2}f(x',x'';l_T^2)\\
       && \frac{\alpha_kk_T^2\vec{e}_1^{(\pm)}\cdot(\vec{k}_T+\vec{l}_T)+
        (k_T^2+(k_T,l_T))\vec{e}_1^{(\pm)}\cdot\vec{k}_T}{(\vec{k}_T+\vec{l}_T)^2}.
\end{eqnarray}
The contribution from Diag.4 is
\begin{eqnarray}
\label{g22}
\nonumber
{\rm Im}{\cal A}^4(\pm,\mp,\pm)&=&{\cal N}'\alpha_k(1+\alpha_k)\int\frac{d^2\vec{l}_T}{(l_T^2)^2}f(x',x'';l_T^2)\\
       && \frac{\alpha_kk_T^2\vec{e}_1^{(\pm)}\cdot(\vec{k}_T-\vec{l}_T)+
        (k_T^2-(k_T,l_T))\vec{e}_1^{(\pm)}\cdot\vec{k}_T}{(\vec{k}_T-\vec{l}_T)^2}.
\end{eqnarray}
The contributions from Diag.5 and Diag.8, to sum up together, are
\begin{eqnarray}
\label{g23}
\nonumber
{\rm Im}{\cal A}^{58}(\pm,\mp,\pm)&=&-{\cal N}'\alpha_k(1+\alpha_k)\int\frac{d^2\vec{l}_T}{(l_T^2)^2}f(x',x'';l_T^2)\\
       && \frac{\alpha_kk_T^2\vec{e}_1^{(\pm)}\cdot(\vec{k}_T-\alpha_k\vec{l}_T)+
        (k_T^2-\alpha_k(k_T,l_T))\vec{e}_1^{(\pm)}\cdot\vec{k}_T}{(\vec{k}_T-\alpha_k\vec{l}_T)^2}.
\end{eqnarray}
The contributions from Diag.6 and Diag.9, to sum up together, are
\begin{eqnarray}
\label{g24}
\nonumber
{\rm Im}{\cal A}^{69}(\pm,\mp,\pm)&=&{\cal N}'(1+\alpha_k)^2\int\frac{d^2\vec{l}_T}{(l_T^2)^2}f(x',x'';l_T^2)\\
       && \frac{\alpha_kk_T^2\vec{e}_1^{(\pm)}\cdot(\vec{k}_T-(1+\alpha_k)\vec{l}_T)+
        (k_T^2-(1+\alpha_k)(k_T,l_T))\vec{e}_1^{(\pm)}\cdot\vec{k}_T}{(\vec{k}_T-(1+\alpha_k)\vec{l}_T)^2}.
\end{eqnarray}
The contributions from other three diagrams (Diag.1, Diag.2 and Diag.7) are
canceled out by each other.
From the above results Eqs.(\ref{g21}-\ref{g24}), we can see that every term
has linear singularity at the limit of $l_T^2\rightarrow 0$ in the integration
of the amplitude over $l_T^2$, while
their total sum is free of the linear singularity.

Following the procedure as we do for the helicity amplitude ${\cal A}(\pm,\mp,\mp)$
in the above, we can approximate the off-diagonal gluon distribution function
$f(x',x'';l_T^2)$ by the usual diagonal differential
gluon distribution function $f_g(x;l_T^2)$.
After integrating over the azimuth angle of $\vec{l}_T$, to sum up all of
Eqs.(\ref{g21}-\ref{g24}), we get the helicity amplitude,
\begin{equation}
\label{gm2}
{\rm Im}{\cal A}(\pm,\mp,\pm)={\cal N}'\pi (1+\alpha_k)^2\vec{e}_1^{(\pm)}\cdot k_T{\cal I},
\end{equation}
where ${\cal I}$ is the same as Eq.(\ref{gi2}) and then Eq.(\ref{gi1}) under
the same approximation.

For the third helicity amplitudes set, ${\rm Im}{\cal A}(\pm,\pm,\mp)$,
the calculations are similar to the calculations of ${\rm Im}{\cal A}(\pm,\mp,\pm)$.
The contribution from Diag.3 is
\begin{eqnarray}
\label{g31}
\nonumber
{\rm Im}{\cal A}^3(\pm,\pm,\mp)&=&-{\cal N}'\alpha_k(1+\alpha_k)\int\frac{d^2\vec{l}_T}{(l_T^2)^2}f(x',x'';l_T^2)\\
       && \frac{(1+\alpha_k)k_T^2\vec{e}_1^{(\pm)}\cdot(\vec{k}_T+\vec{l}_T)-
        (k_T^2+(k_T,l_T))\vec{e}_1^{(\pm)}\cdot\vec{k}_T}{(\vec{k}_T+\vec{l}_T)^2}.
\end{eqnarray}
The contribution from Diag.4 is
\begin{eqnarray}
\label{g32}
\nonumber
{\rm Im}{\cal A}^4(\pm,\pm,\mp)&=&{\cal N}'\alpha_k^2\int\frac{d^2\vec{l}_T}{(l_T^2)^2}f(x',x'';l_T^2)\\
       && \frac{(1+\alpha_k)k_T^2\vec{e}_1^{(\pm)}\cdot(\vec{k}_T-\vec{l}_T)-
        (k_T^2-(k_T,l_T))\vec{e}_1^{(\pm)}\cdot\vec{k}_T}{(\vec{k}_T-\vec{l}_T)^2}.
\end{eqnarray}
The contributions from Diag.5 and Diag.8, to sum up together, are
\begin{eqnarray}
\label{g33}
\nonumber
{\rm Im}{\cal A}^{58}(\pm,\pm,\mp)&=&-{\cal N}'\alpha_k^2\int\frac{d^2\vec{l}_T}{(l_T^2)^2}f(x',x'';l_T^2)\\
       && \frac{(1+\alpha_k)k_T^2\vec{e}_1^{(\pm)}\cdot(\vec{k}_T-\alpha_k\vec{l}_T)-
        (k_T^2-\alpha_k(k_T,l_T))\vec{e}_1^{(\pm)}\cdot\vec{k}_T}{(\vec{k}_T-\alpha_k\vec{l}_T)^2}.
\end{eqnarray}
The contributions from Diag.6 and Diag.9, to sum up together, are
\begin{eqnarray}
\label{g34}
\nonumber
{\rm Im}{\cal A}^{69}(\pm,\pm,\mp)&=&{\cal N}'\alpha_k(1+\alpha_k)\int\frac{d^2\vec{l}_T}{(l_T^2)^2}f(x',x'';l_T^2)\\
       && \frac{(1+\alpha_k)k_T^2\vec{e}_1^{(\pm)}\cdot(\vec{k}_T-(1+\alpha_k)\vec{l}_T)-
        (k_T^2-(1+\alpha_k)(k_T,l_T))\vec{e}_1^{(\pm)}\cdot\vec{k}_T}{(\vec{k}_T-(1+\alpha_k)\vec{l}_T)^2}.
\end{eqnarray}
And also, we find that the contributions from other three diagrams (Diag.1, Diag.2 and Diag.7) are
canceled out by each other, and the total sum of 
Eqs.(\ref{g31}-\ref{g34}) is free of the linear singularity.
If we approximate the off-diagonal gluon distribution function
$f(x',x'';l_T^2)$ by the usual diagonal differential
gluon distribution function $f_g(x;l_T^2)$,
and integrate over the azimuth angle of $\vec{l}_T$, their sum will lead to
a similar result as in Eq.(\ref{gm2}),
\begin{equation}
\label{gm3}
{\rm Im}{\cal A}(\pm,\pm,\mp)={\cal N}'\pi \alpha_k^2\vec{e}_1^{(\pm)}\cdot k_T{\cal I}.
\end{equation}

By summing up all of the helicity amplitudes Eqs.(\ref{gm1}), (\ref{gm2}) and
{\ref{gm3}), we will get the amplitude squared for 
the partonic process $gp\rightarrow ggp$, after averaging
over the spin and color degrees of freedom,
\begin{equation}
\overline{|{\cal A}|}^2=\frac{27\pi^2\alpha_s^3(4\pi)^3}{16}\frac{s^2}{k_t^2}(1-\frac{k_T^2}{M_X^2})^2|{\cal I}|^2.
\end{equation}
And the cross section for the partonic process $gp\rightarrow ggp$ is
\begin{equation}
\label{gxsp}
\frac{d\hat\sigma(gp\rightarrow ggp)}{dt}|_{t=0}=\int_{M_X^4>4k_T^2}dM_X^2dk_T^2
        \frac{27\alpha_s^3\pi^2}{32M_X^2k_T^2}(1-\frac{k_T^2}{M_X^2})^2|{\cal I}|^2\frac{1}{\sqrt{1-\frac{4k_T^2}{M_X^2}}},
\end{equation}
Following the same argument in the last subsection for the calculations of the partonic process
$qp\rightarrow qgp$, 
we see that the dominant contribution of the integration over $M_X^2$
comes from the region of $M_X^2\sim 4k_T^2$, where
the differential gluon distribution
function $f_g(x;Q^2)$ of the three terms in the integration ${\cal I}$
can approximately take their values
at the same scale of $Q^2=k_T^2$.
That is, the integration ${\cal I}$ is then simplified to
\begin{equation}
\label{i22}
{\cal I}=\frac{1}{k_T^2}(-3\alpha_k(1+\alpha_k))f_g(x;k_T^2)
=\frac{1}{k_T^2}\frac{3k_T^2}{M_X^2}f_g(x;k_T^2)=\frac{3}{M_X^2}f_g(x;k_T^2).
\end{equation}

\section{Numerical results}

In this section we study the numerical behavior of the diffractive gluon jet
production at the Fermilab Tevatron. We will study the $p_T$ distribution
and $x_1$ distribution of the cross section. We will also compare the gluon
jet production with the quark jet production which has been calculated in
\cite{charm,quark}. A more thorough phenomenological study, including a comparison
to currently available data at Tevatron on the diffractive dijet production
rate, will be presented elsewhere.

Provided with the cross section formulas for the partonic processes
$qp\rightarrow qg p$ (\ref{xsp}) and $gp\rightarrow ggp$ (\ref{gxsp}),
we can calculate the cross
section of the diffractive gluon jet production
at hadron level.
However, as mentioned above, there exists nonfactorization effect caused by
the spectator interactions in the hard 
diffractive processes in hadron collisions.
Here, we use a suppression factor ${\cal F}_S$ to describe this
nonfactorization effect in the hard diffractive processes at hadron
colliders\cite{soper,survive}.
At the Tevatron,
the value of ${\cal F}_S$ may be as small as ${\cal F}_S\approx 0.1$\cite{soper,tev}.
That is to say, the total cross section of the diffractive processes
at the Tevatron may be reduced down by an order of magnitude due to
this nonfactorization effect.
In the following numerical calculations, we adopt this suppression factor value
to evaluate the diffractive production rate.

In our calculations, the scales for the parton distribution functions and
the running coupling constant
are both set to be $Q^2=k_T^2$. For the parton distribution functions,
we choose the GRV NLO set \cite{grv}.

In Fig.4, we plot the differential cross section $d\sigma/dt|_{t=0}$
as a function of the lower bound of the transverse momentum of the gluon
jet, $k_{T{\rm min}}$. This figure shows that the cross section is
sensitive to the transverse momentum cut $k_{T{\rm min}}$.
We plot separately the contributions from the two subprocesses,
$qp\rightarrow qgp$ and $gp\rightarrow ggp$.
By comparison, we also plot the cross section of 
the diffractive light quark jet
production calculated in\cite{quark}.
The three curves in this figure show that the contribution from the subprocess
$gp\rightarrow ggp$ is two orders of magnitude larger than that from
the subprocess $qp\rightarrow qgp$ for the diffractive gluon jet production,
and the light quark jet production rate is in the same order with that of the
subprocess $qp\rightarrow qgp$.
This indicates that the diffractive dijet production at hadron colliders
dominantly comes from the subprocess $gp\rightarrow ggp$ in the two-gluon
exchange model.

In Fig.5, we plot the differential cross section $d\sigma/dt|_{t=0}$ as
a function of the lower bound of the momentum fraction of the proton
carried by the incident gluon $x_{1{\rm min}}$,
where we set $k_{T{\rm min}}=5~GeV$.
Fig.5(a) is for the contribution from the subprocess
$qp\rightarrow qgp$, and Fig.5(b) is
from the subprocess $gp\rightarrow ggp$.
These two figures show that the dominant
contribution comes from the region of $x_1\sim 10^{-2}-10^{-1}$ for the
subprocess $gp\rightarrow ggp$, and $x_1>10^{-1}$ for the subprocess
$qp\rightarrow qgp$.
These properties are similar to those of the diffractive charm jet
and $W$ boson productions calculated in\cite{charm,dy}.

\section{Conclusions}

In this paper, we have calculated the diffractive gluon jet production
at hadron colliders in perturbative QCD by using the two-gluon exchange model.
We find that the production cross section is related to the squared of the
differential gluon distribution function $\partial G(x;Q^2)/\partial ln Q^2$
at the scale of $Q^2\sim k_T^2$, where $k_T$ is the transverse momentum of
the final state gluon jet.
We have also compared the production rate of the gluon jet in the
diffractive processes with those of the light quark jet and heavy quark jet productions,
and found that the production rates of these processes
are in the same order of magnitude.

As we know, the large transverse momentum dijet production in the diffractive
processes at hadron colliders is important to study the diffractive mechanism and
the nature of the Pomeron. The CDF collaboration at the Fermilab Tevatron
have reported some results on this process\cite{tev}.
Up to now, we have calculated all of the dijet production subprocesses
in the diffractive processes at hadron colliders, including $gp\rightarrow
q\bar qp$, $qp\rightarrow qgp$ and $gp\rightarrow ggp$ processes.
In a forthcoming paper, we will compare the available data on the diffractive
dijet production cross section at the Tevatron\cite{tev} to the predictions of
our model to test the validity of perturbative QCD description of the diffractive
processes at hadron colliders.

\acknowledgments
This work was supported in part by the National Natural Science Foundation
of China, the State Education Commission of China, and the State
Commission of Science and Technology of China.


\newpage
\vskip 10mm
\centerline{\bf \large Figure Captions}
\vskip 1cm
\noindent
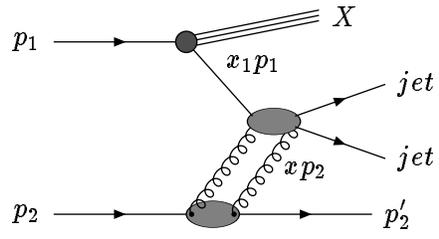
Fig.1. Sketch diagram for the diffractive dijet production at hadron colliders
in perturbative QCD. The final state jet lines represent the outgoing quark
or gluon lines, and the incident parton from the upper proton (labeled by $x_1p_1$)
can be quark or gluon correspondingly.

\noindent
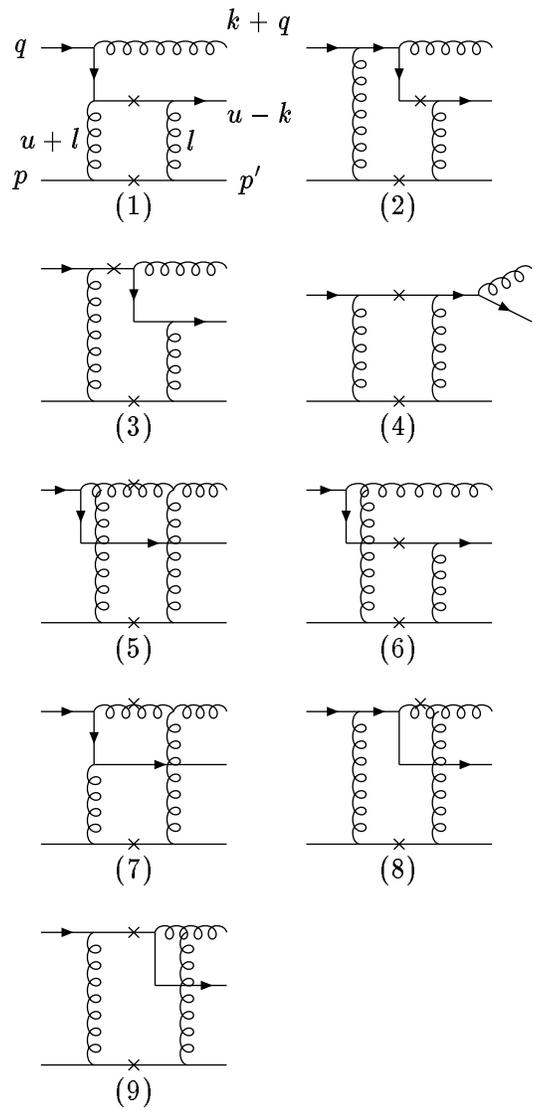
Fig.2. The lowest order perturbative QCD diagrams for partonic process
$qp\rightarrow qg p$.

\noindent
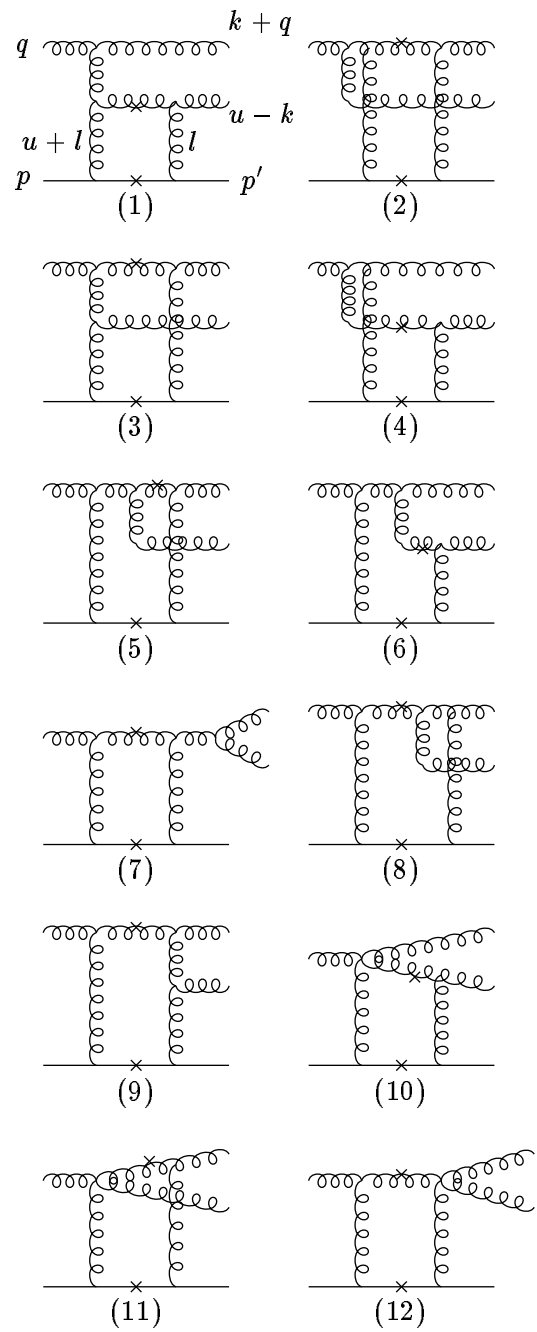
Fig.3. The lowest order perturbative QCD diagrams for partonic process
$gp\rightarrow gg p$.

\noindent
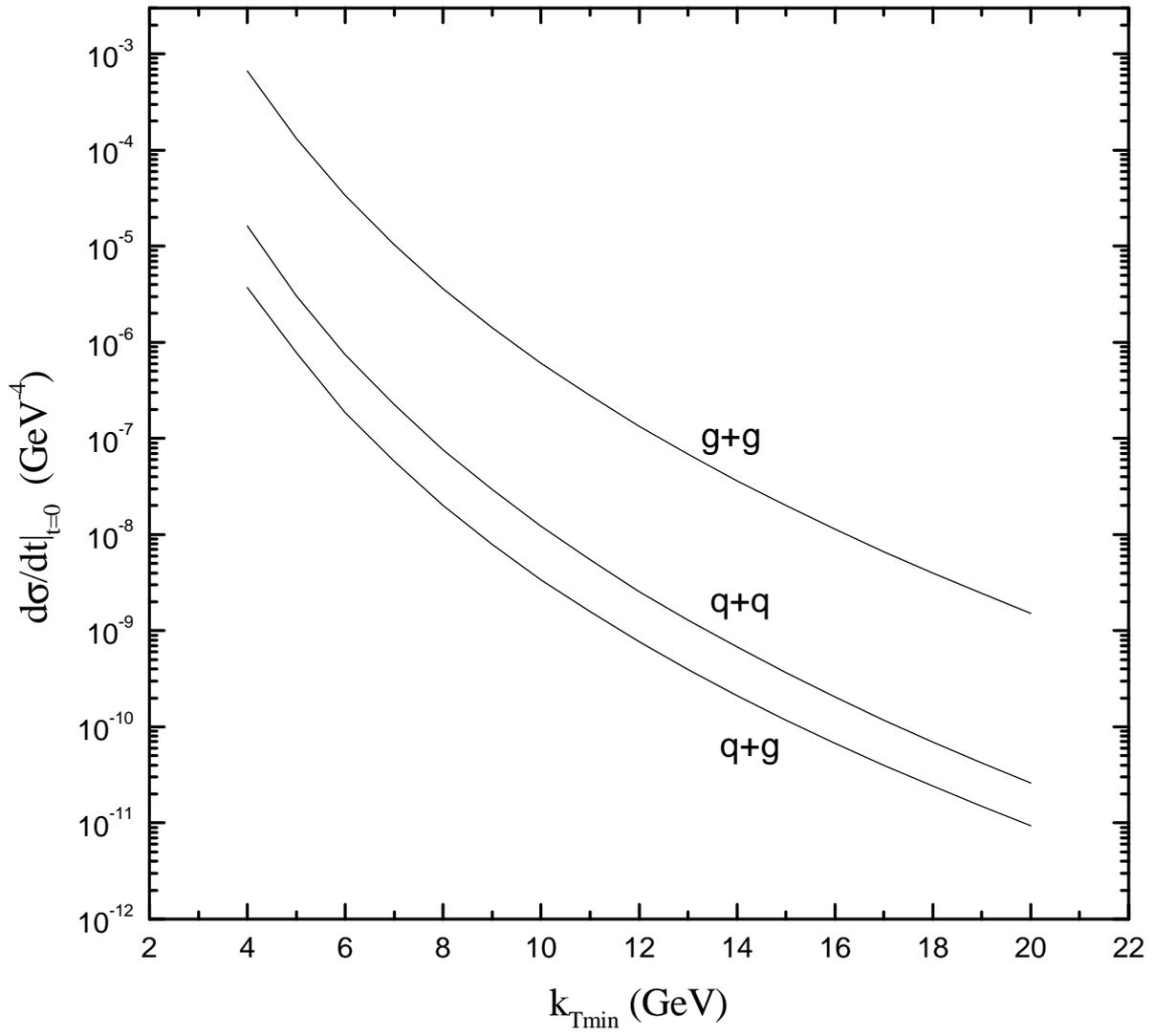
Fig.4. The differential cross section $d\sigma/dt|_{t=0}$ for the gluon jet
production in the diffractive processes as a function
of $k_{T{\rm min}}$ at the Fermilab Tevatron,
where $k_{T{\rm min}}$ is the lower bound of the transverse momentum of the
out going gluon jet.

\noindent
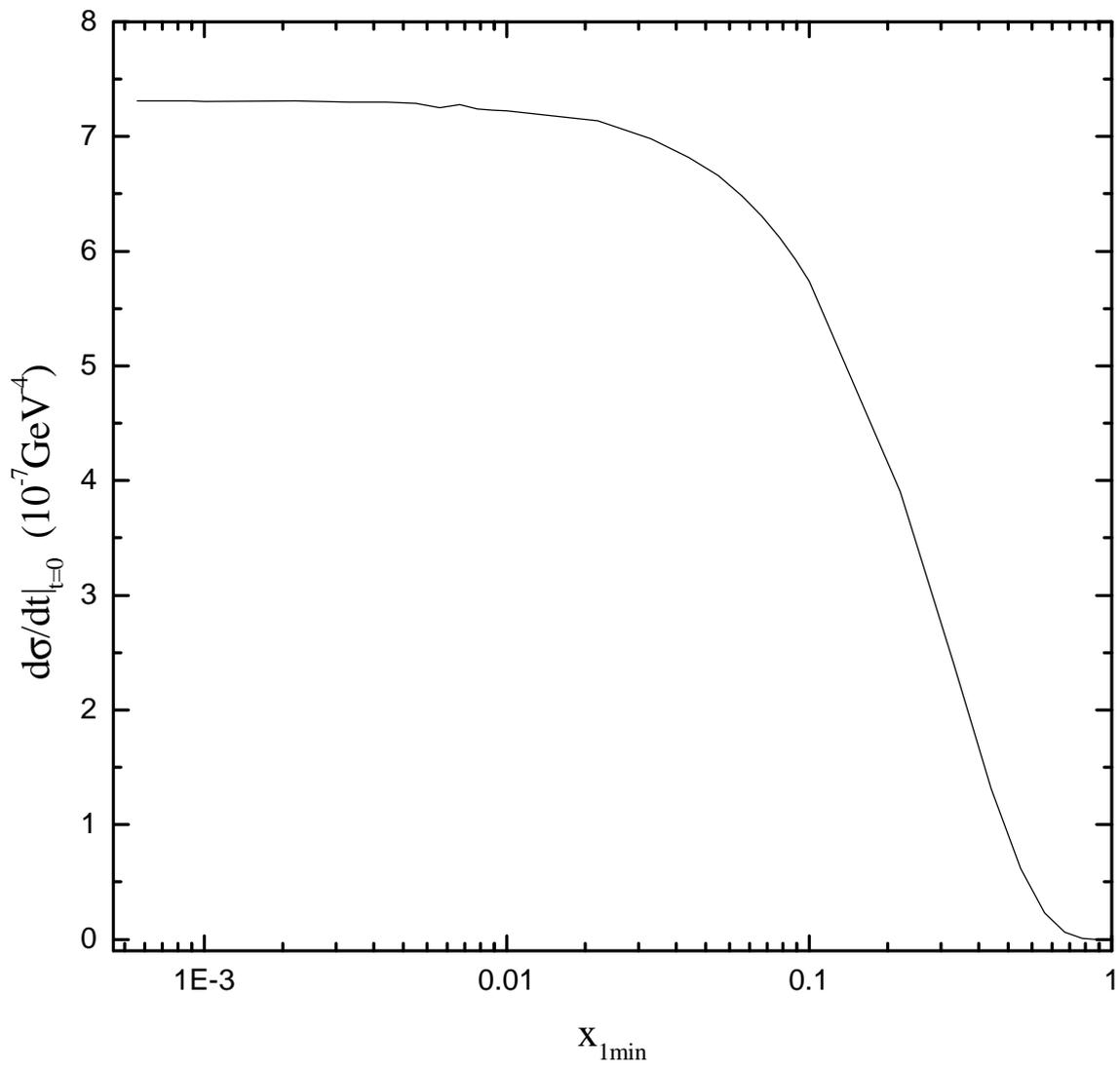
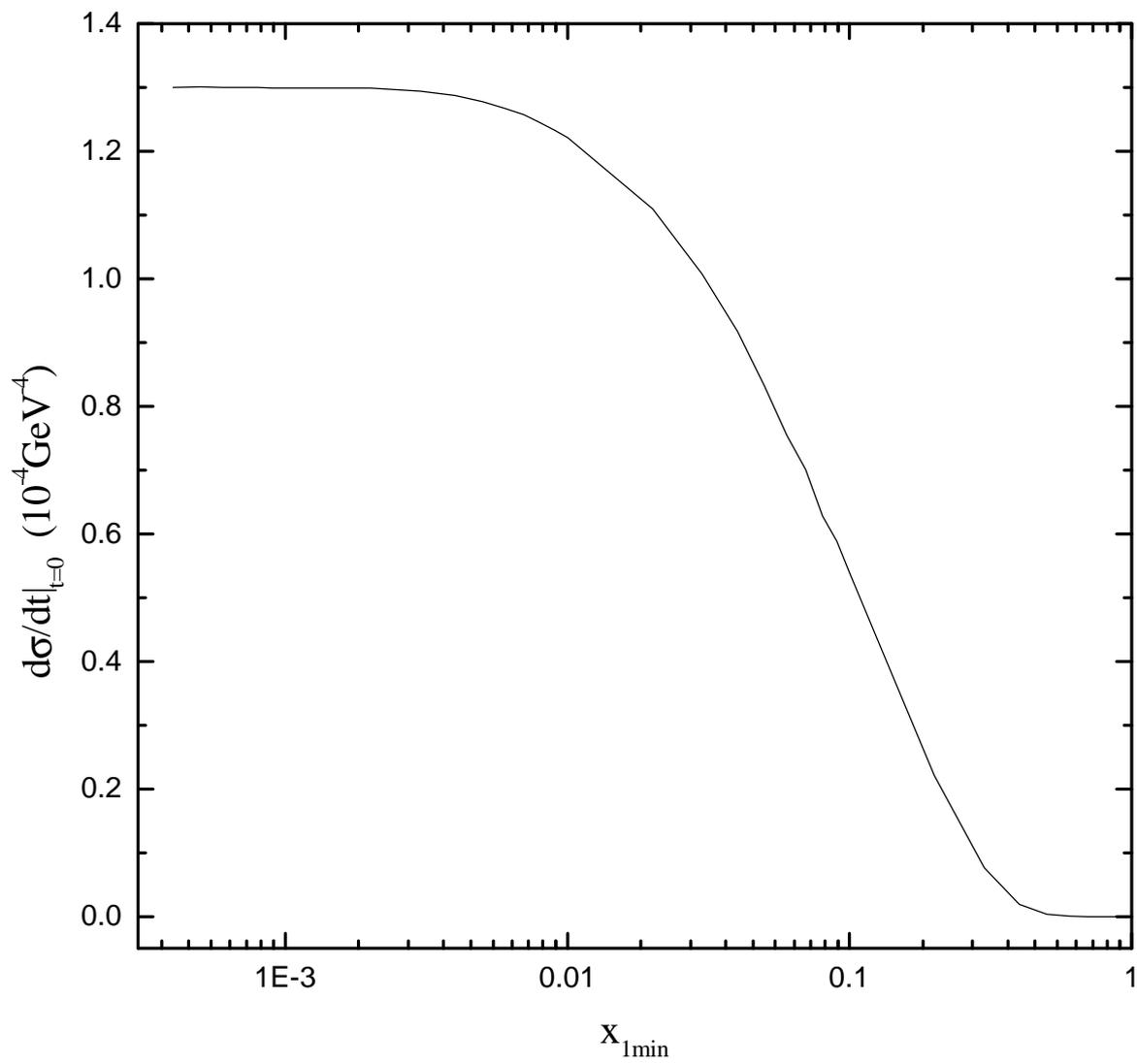
Fig.5. The differential cross section $d\sigma/dt|_{t=0}$ for the gluon jet
production
as a function of $x_{1{\rm min}}$, where $x_{1{\rm min}}$ is the lower
bound of $x_1$ in the integration of the cross section.
(a) is for the contribution from the subprocess
$qp\rightarrow qgp$, and (b) is
from the subprocess $gp\rightarrow ggp$.

\end{document}